\documentclass[conference]{IEEEtran}
\IEEEoverridecommandlockouts

\usepackage{pgfplots}
\usepackage{balance}
\usepackage{amsthm}
\usepackage[inline]{enumitem}
\usepackage{color}
\usepackage{float}
\usepackage[cmex10]{amsmath}
\usepackage{amssymb,amsthm}
\usepackage{mathrsfs}
\usepackage{amsbsy}
\usepackage{multirow}
\usepackage{graphicx}
\usepackage{amsfonts}
\usepackage{xcolor}
\usepackage{booktabs}
 \usepackage{graphicx}
 \if CLASSOPTIONcompsoc
 \usepackage[caption=false, font=normalsize, labelfont=sf, textfont=sf] {subfig}
 \else
 \usepackage[caption=false, font=footnotesize]{subfig}
\fi

\usepackage[numbers]{natbib}

\makeatletter
\def\testclr#1#{\@testclr{#1}}
\def\@testclr#1#2{{\fboxsep\z@\fbox{\colorbox#1{#2}{\phantom{XX}}}}}

\makeatother

\usepackage{glossaries}
\newacronym{cdf}{CDF}{cumulative distribution function}
\newacronym{nlos}{NLOS}{non-line-of-sight}
\newacronym{los}{LOS}{line-of-sight}
\newacronym{csi}{CSI}{channel state information}
\newacronym{cir}{CIR}{channel impulse response}
\newacronym{agc}{AGC}{automatic gain control}
\newacronym{boi}{BoI}{band-of-interest}
\newacronym{psd}{PSD}{power spectral density}
\newacronym{dft}{DFT}{discrete Fourier transform}

\newacronym{cfr}{CFR}{channel frequency response}
\newacronym{uwb}{UWB}{ultra-wideband}
\newacronym{kkt}{KKT}{Karush-Kuhn-Tucker}
\newacronym{cnc}{CNC}{computer numerical control}
\newacronym{rp}{RP}{recurrence plot}
\newacronym{iot}{IoT}{Internet of Things}

\begin{document}

\title{Band-of-Interest-based Channel Impulse Response Fusion for Breathing Rate Estimation with UWB
\thanks{This research has been kindly supported by the Swiss National Science Foundation under Grant-ID 182621.}
}

\author{\IEEEauthorblockN{Sitian Li}
\IEEEauthorblockA{\textit{Telecommunication Circuits Laboratory} \\
\textit{\'{E}cole polytechnique f\'{e}d\'{e}rale de Lausanne}\\
Switzerland \\
sitian.li@epfl.ch}
\and
\IEEEauthorblockN{Alexios Balatsoukas-Stimming}
\IEEEauthorblockA{\textit{Department of Electrical Engineering} \\
\textit{Eindhoven University of Technology}\\
the Netherlands \\
a.k.balatsoukas.stimming@tue.nl}
\and
\IEEEauthorblockN{Andreas Burg}
\IEEEauthorblockA{\textit{Telecommunication Circuits Laboratory} \\
\textit{\'{E}cole polytechnique f\'{e}d\'{e}rale de Lausanne}\\
Switzerland\\
andreas.burg@epfl.ch}
}

\maketitle

\begin{abstract}
The channel impulse response (CIR) obtained from the channel estimation step of various wireless systems is a widely used source of information in wireless sensing.  
Breathing rate is one of the important vital signs that can be retrieved from the CIR.
Recently, there have been various works that extract the breathing rate from one carefully selected CIR delay bin that contains the breathing information.
However, it has also been shown that the accuracy of this estimation is very sensitive to the measurement scenario, e.g., if there is any obstacle between the transceivers and the target, the position of the target, and the orientation of the target, since only one CIR delay bin does not contain a sufficient periodic component to retrieve the breathing rate.
We focus on such scenarios and propose a CIR delay bin fusion method to merge several CIR bins to achieve a more accurate and reliable breathing rate estimate.
We take measurements and showcase the advantages of the proposed method across scenarios.

\end{abstract}

\begin{IEEEkeywords}
Breathing rate estimation, Channel impulse response, Ultra-wideband
\end{IEEEkeywords}

\IEEEpeerreviewmaketitle

\section{Introduction}
\label{sec:intro}
Detecting and monitoring vital signs using \gls{iot} devices have become trending topics in the last decade~\cite{jara_knowledge_2012, gupta_healthcare_2015, wu_internet--things_2018} and have a wide range of applications such as elderly care, sleep monitoring, and home surveillance.
Various sensors are integrated into different wearable devices~\cite{iqbal_advances_2021, ganchev_healthcare_2019} and home electronics to collect vital sign information.
Apart from built-in sensors, wireless communication signals can also be used to extract environment information.
Nowadays with the emerging wireless communications technologies, wireless communication signals are ubiquitous in the environment.
In fact, communication signals show considerable potential to be utilized as a source for different types of information, including vital signs.
Breathing rate is one of the most important vital signs and an early and extremely good indicator of physiological conditions~\cite{nicolo_importance_2020}.
A normal and healthy breathing rate is stable and in the range of $0.1$\,Hz to $0.5$\,Hz~\cite{benchetrit_breathing_2000}.
Radar systems have been used for breathing rate estimation before wireless communication signals, and the breathing rate is calculated from the round-trip time of the radar signal that is reflected on the human chest~\cite{droitcour_signal--noise_2009, costanzo_software-defined_2019}.

Apart from radar, there is a more convenient solution using WiFi or \gls{uwb}  transceivers, from which a more detailed channel characteristic, namely the \gls{cir}, can be measured.
The \gls{cir} is a complex-valued vector that represents the magnitude and phase of the signal paths across multiple different delays.
The wireless signal that is reflected by the human chest shows a periodic feature across time~\cite{wang_phasebeat_2017}.
Multiple \gls{cir} snapshots must be recorded over time to extract breathing rate information.
However, since each \gls{cir} snapshot contains multiple delay bins, an effective dimension-reduction method needs to be developed.
The most intuitive method is to select one \gls{cir} delay bin and observe it over time. 
There are several selection criteria, such as the periodicity~\cite{dou_full_2021} calculated from the \gls{rp}~\cite{eckmann_recurrence_1987} and the signal energy in the frequencies of breathing rate (\gls{boi})~\cite{li_complexbeat_2021}.
Other works assume prior knowledge of the position of the target and select the bins of interest~\cite{goldfine_respiratory_2020, li_complexbeat_2021}.
A \gls{dft} can be applied to each \gls{cir} snapshot to obtain the \gls{cfr}, which is also referred to as \gls{csi}.
The \gls{csi} is a complex-valued vector whose entries represent the amplitude and phase of different subcarriers, which are different linear combinations of the delay bins in the \gls{cir} based on the \gls{dft} matrix.
These can also be used to retrieve the  breathing rate.
The mean absolute deviation in the phase of the \gls{csi}~\cite{wang_phasebeat_2017}, the variance of the complex-valued \gls{csi}~\cite{liu_tracking_2015}, and the periodicity of \gls{csi}~\cite{liu_wi-sleep_2014} have been proposed as selection criteria for choosing one subcarrier in the \gls{csi} for breathing rate estimation.

Unfortunately, there is a drawback to all these selection methods.
When the breathing components are spread across more than one \gls{cir} bin or \gls{csi} subcarrier, selecting only one bin or subcarrier leads to a loss of information and impacts the performance.
Most previous works assume that the target is in the \gls{los} of both the transmitter and the receiver.
In such cases, the \gls{cir} bin that corresponds to a direct reflection on the human chest contains most of the relevant breathing period information.
For the \gls{nlos} case, the signal on the direct reflection path can be attenuated by obstacles.
Moreover, based on our observation and conclusions from previous work~\cite{wang_human_2016}, even for the \gls{los} case, the breathing rate estimation performance is very sensitive to the location and the orientation of the target.
In practice, assuming the location, orientation of the targets or the absence of obstacles in front of the target restricts valid locations of the target, or constrains the choice of the mounting position of the measurement devices, i.e., the transmitter and the receiver in order to provide a reasonably extensive coverage area.
Especially when the transmitter and the receiver are embedded in personal electronic devices such as tablets and cell phones, it is not convenient for users to put the device in a deliberated position for breathing rate estimation.
In a more complex environment, where there are more obstacles, especially between the target and the transceivers, and the location and the orientation of the target are unknown, it is hard to determine which \gls{cir} bin contains the most breathing component and due to multiple reflections, the periodic components in each individual bin may be very small.

\subsection*{Contributions}
In this work, we propose to focus on and consider all \gls{cir} bins that contain periodic components that are in the range of human breathing rate.
Only the static or noisy bins in the \gls{cir} are discarded during breathing rate estimation.
This is achieved by combining multiple \gls{cir} bins based on their energy in the \gls{boi}.
To this end, we formulate a \gls{boi}-based fusion method to linearly combine multiple \gls{cir} bins such that the combined signal power in the \gls{boi} is maximized to improve the breathing rate estimation performance.

Moreover, we build up a breathing rate measurement setup with \gls{uwb} transceivers.
To estimate the breathing rate, we obtain the calibrated \gls{cir} and merge the \gls{cir} bins based on our \gls{boi} fusion method.
We use a moving \gls{cnc} machine as a breathing emulator to simulate the movement of the human chest located at different distances and orientations from the \gls{uwb} transceivers.
We also examine both cases where the breathing emulator is and is not in the \gls{los} of the transceivers and compare the estimation performance with the \gls{boi} selection method implemented in~\cite{li_complexbeat_2021}.
We demonstrate that our proposed method shows especially better performance compared to the \gls{boi} selection method \cite{li_complexbeat_2021} in the \gls{nlos} case and when the breathing movement is small.
Finally, we validate the results with real human breathing experiments.

\section{Background}
In this section, we introduce the background for breathing rate estimation using the \gls{cir}.
We first introduce the structure of the \gls{cir} and the employed \gls{cir} calibration method.
Then we discuss how the \gls{cir} is influenced by breathing movement, and introduce the mathematical expression for the \gls{boi} for breathing rate estimation. 
In the end, we introduce the \gls{psd}-based breathing rate estimation method.
\subsection{Channel Impulse Response (CIR) and CIR calibration}
\label{sec:calib}
The ground truth \gls{cir} can be written as~\cite{li_device-free_2022}
\begin{align}
h\left(\tau\right) = \sum_{p=1}^{P} A_p g\left(\tau - \tau_p\right) e^{-j 2 \pi f_c \tau_p},
\label{equ:pre_cir_multi_path}
\end{align}
where $g\left(\tau\right) = \frac{\sin\left(\pi B \tau\right)}{\pi B \tau}$ is the pulse shape function, $P$ is the number of multipath components, $A_p$ and $\tau_p$ are the attenuation and delay of path $p$, respectively, $B$ is the bandwidth and $f_c$ is the carrier frequency.
A coherent \gls{uwb} receiver determines the exact \gls{cir} between the transmitter and the receiver by a preamble sequence that has a perfect periodic auto-correlation.
The \gls{cir} we obtain from the \gls{uwb} receiver is a sampled \gls{cir} vector, $\mathbf{h}$, whose entries are $h_n = h(\frac{n}{B}) $.

There is usually a delay offset and an amplitude distortion in the estimated \gls{cir} because of the non-synchronized transmitter and the receiver and the \gls{agc} in the receiver, respectively.
We calibrate the \gls{cir} delay offset by using the method described in~\cite{li_device-free_2022}.
To calibrate the amplitude distortion, we use the same amplitude calibration method as proposed in~\cite{li_complexbeat_2021}.

\subsection{Impact of Breathing Movement on CIR with \gls{nlos}}
In a complex environment with numerous obstacles, a direct reflection path on the human chest can be attenuated. 
However, all the transmission paths that have been reflected at least once on the target contain the periodic component that lies in the \gls{boi}.
As shown in Fig.~\ref{fig:room_ray}, there are multiple paths reflected on different sets of objects including the target. 
Different from the direct path with only one reflection which has a distinct delay that is related to the distance between the transmitter and receiver (referred to as the transceivers) and the target, the other paths with multiple reflections have various delay values.
Therefore, multiple delay bins in the \gls{cir} may also contain valuable breathing information. 
\subsection{Band-of-Interest for Breathing Rate}
The energy in the \gls{boi} for breathing is defined as the sum of the entries of the \gls{psd} of a snapshot sequence that are within the possible breathing rate range.
\begin{align}
P_{I} &= {\sum\nolimits_{f_r^l \leq |f| \leq f_r^h} {{{\left| {{P}[f]} \right|}^2}} } \\
&= \left(\mathbf{F}_I^H  \mathbf{x}\right)^H \mathbf{F}_I^H \mathbf{x} \label{equ:boi_equ},
\end{align}
where $\mathbf{F}_I$ denotes rows in the \gls{dft} matrix corresponding to the frequencies between $f_r^l$ and $f_r^h$, where $f_r^l$ and $f_r^h$ are the lower and upper bounds of human breathing rate, respectively.
In the work of~\cite{li_complexbeat_2021}, the \gls{cir} bin that contains the most energy in the \gls{boi} is selected for breathing rate estimation,  which shows good performance compared to other methods in~\cite{li_complexbeat_2021}.
We use this method as the baseline (\gls{boi} selection), and the selected \gls{cir} bin is referred to as $\mathbf{x}_k$ in the following.

\subsection{Breathing Rate Estimation}
In order to obtain the exact breathing rate in the \gls{boi}, we need to apply a \gls{dft} to the snapshot sequence $\mathbf{x}_k$ to obtain the \gls{psd} of the snapshot sequence and detect the peak value. This method is referred to as the \gls{psd}-detect method in~\cite{hillyard_comparing_2018, li_complexbeat_2021}.

\section{System Structure}
\begin{figure}[t!] 
  \centering
  \includegraphics[width=0.97\linewidth]{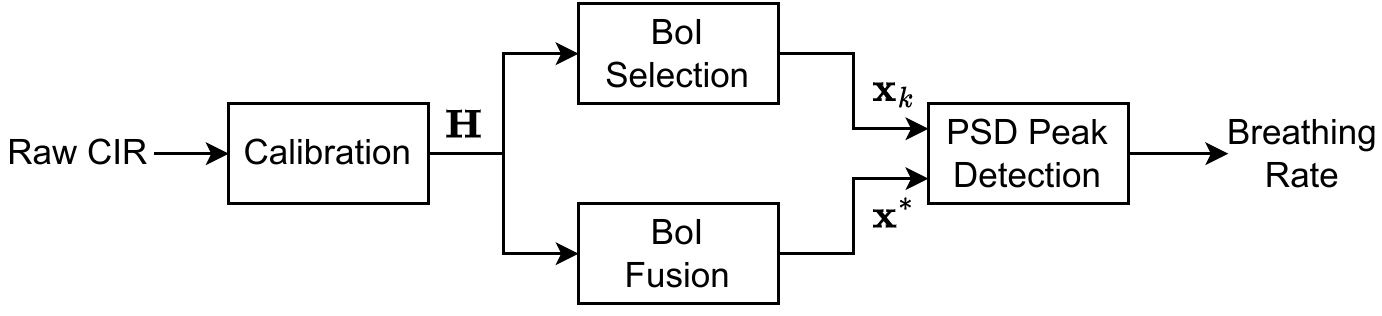}
  \caption{Block diagram of data processing pipelines for the BoI selection and BoI fusion methods.}
  \label{fig:block_diagram} 
\end{figure}

The objective of this work is to estimate the breathing rate of humans in different positions and orientations by using \gls{uwb} transceivers.
In order to deal with the cases where the periodic component is small but presents in multiple different \gls{cir} bins, such as when the person is not in the \gls{los} of the transceivers and when the person is not facing the transceivers, we introduce the \gls{boi} fusion method to combine multiple relevant \gls{cir} bins.
The corresponding system structure is shown in Fig.~\ref{fig:block_diagram}.
We implement both the \gls{boi} selection method~\cite{li_complexbeat_2021} and our \gls{boi} fusion method to compare the performance.
In this section, we introduce the calibration method we implement and the \gls{boi} fusion method we propose.
\subsection{\Gls{cir} Calibration with Transmission Line}
\begin{figure}[t!] 
  \centering
  \subfloat[Raw CIR amplitude\label{fig:cir_raw}]{%
        \includegraphics[width=0.57\linewidth]{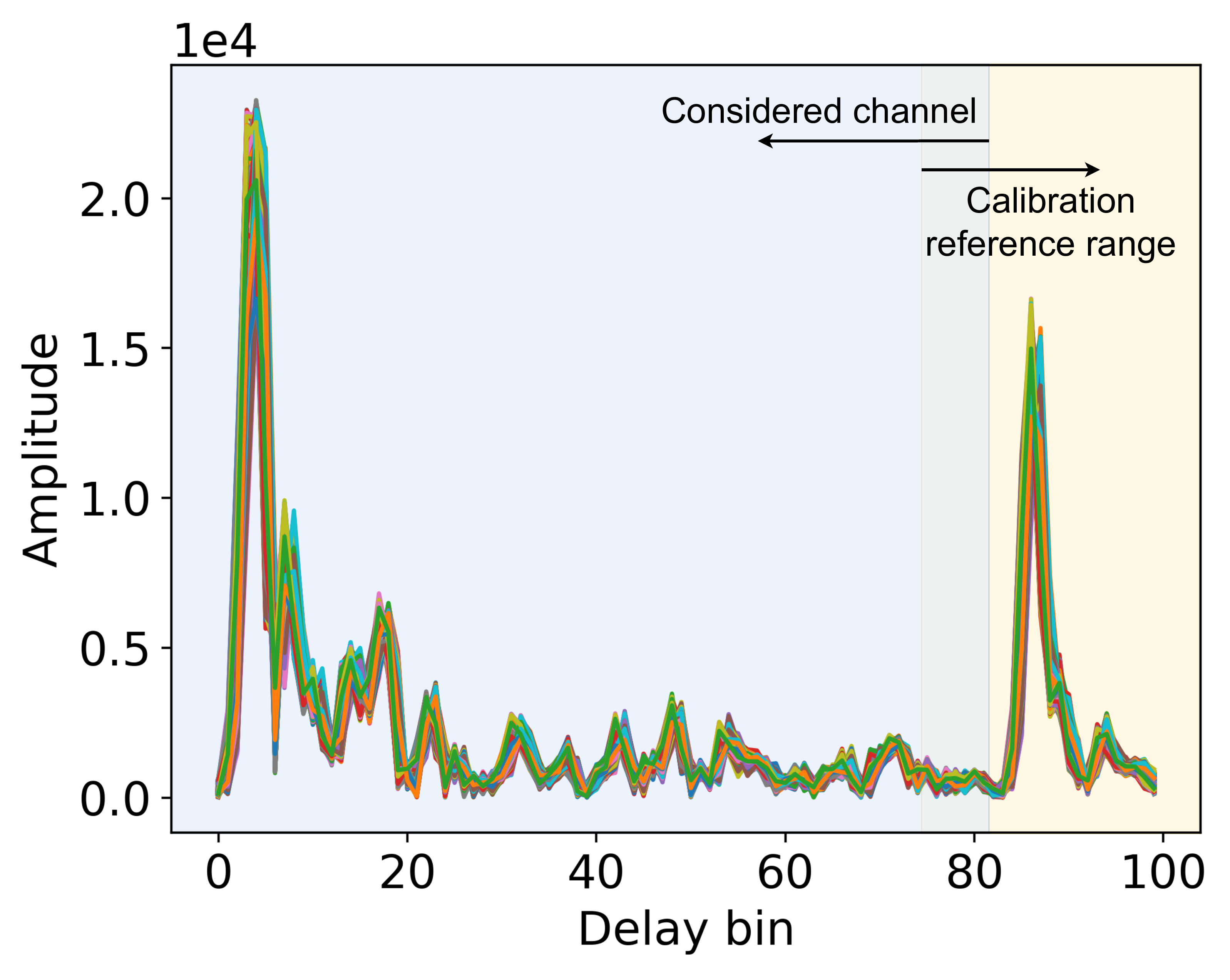}}
    \hfill
  \subfloat[Calibrated CIR amplitude\label{fig:cir_calib}]{%
        \includegraphics[width=0.57\linewidth]{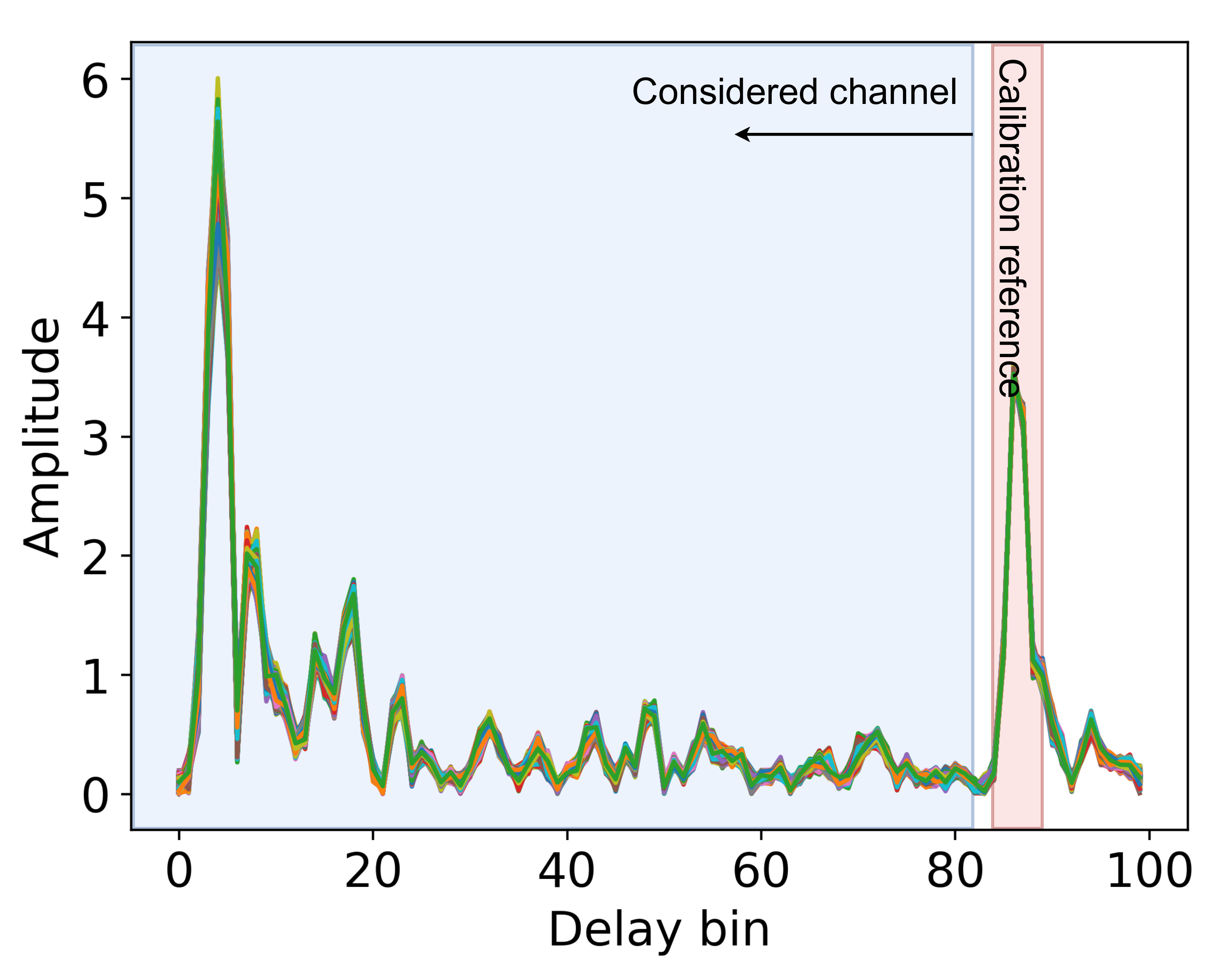}}
  \caption{CIR amplitude obtained before and after calibration}
  \label{fig:calibration} 
\end{figure}

In order to compare the performance  of the \gls{boi} fusion method with the baseline method, i.e., the \gls{boi} selection method, we need to first calibrate the \gls{cir} with high precision.
The phase and amplitude are calibrated as described in~\ref{sec:calib}. 
Instead of assuming a static \gls{los} path between the transmitter and the receiver, we insert a transmission line between the transmitter and the receiver as the reference path for calibration.
We use power splitters on the transmitter and the receiver such that part of the RF signal is transmitted through the transmission line with a fixed delay with an attenuator as a known reference for calibration.
The transmission line should be longer than the wireless channel we take into consideration, and in our setup, we use a transmission line that is about~$25$\,m long.
The \gls{uwb} receiver synchronizes with the shortest path to the transmitter, which is the first peak in Fig.~\ref{fig:calibration}.
As mentioned in Section~\ref{sec:calib}, there is always a delay offset and amplitude difference between the snapshots.
Therefore, as shown in Fig.~\ref{fig:cir_raw}, before calibration, the \gls{cir}s are shifted with different delay offsets and the amplitude for different snapshots varies.
To calibrate the delay offset, we need to find the static reference signal peak position in the \gls{cir}, which corresponds to the signal transmitted through the transmission line.
Knowing that the distance between the transmitter and the receiver is $0.5$\,m and the distance between adjacent bins in the \gls{cir} that we obtained is~$1$\,ns, we can expect the peak from the transmission line to be about $\frac{(25-0.5)\text{m}}{c\times 1\text{\,ns}} \approx 82$~bins after the first peak in the \gls{cir}, where $c$ is the speed of light.
Thus we pick the bins after bin~$75$ as the potential calibration reference.
By finding the bin with the maximum amplitude within the calibration reference range and its four adjacent bins and using these five bins as the reference, the synchronized delay offset can be faithfully removed.
Then we calibrate the amplitude based on the \gls{cir} with calibrated delay.
We search the maximum bins again within the reference range, and use the maximum bin and its four adjacent bins to calibrate the amplitude such that the energy contained in the reference bins is the same for all snapshots.
Fig.~\ref{fig:cir_calib} shows the calibrated \gls{cir}s, where the random delay offset and amplitude difference are removed.
In the following, the calibrated \gls{cir} is denoted by a complex-valued vector $\mathbf{h}$ and a snapshot sequence of $\mathbf{h}$ is denoted by a complex-valued matrix $\mathbf{H}$, where each row represents one $\mathbf{h}$ at a given time instance.

\begin{figure}[t!] 
  \centering
  \subfloat[Room layout\label{fig:lab_raw}]{%
        \includegraphics[width=0.4\linewidth]{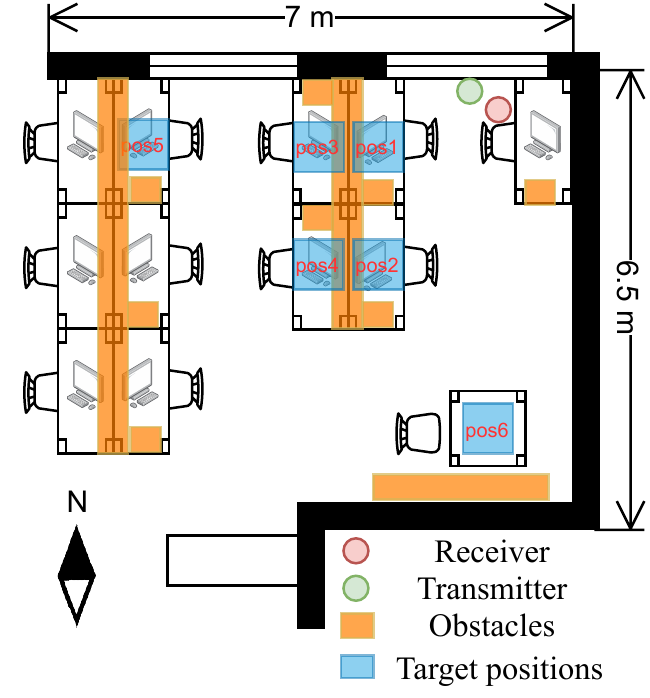}}
    \hfill
  \subfloat[Transmission paths with obstabcles\label{fig:room_ray}]{%
        \includegraphics[width=0.53\linewidth]{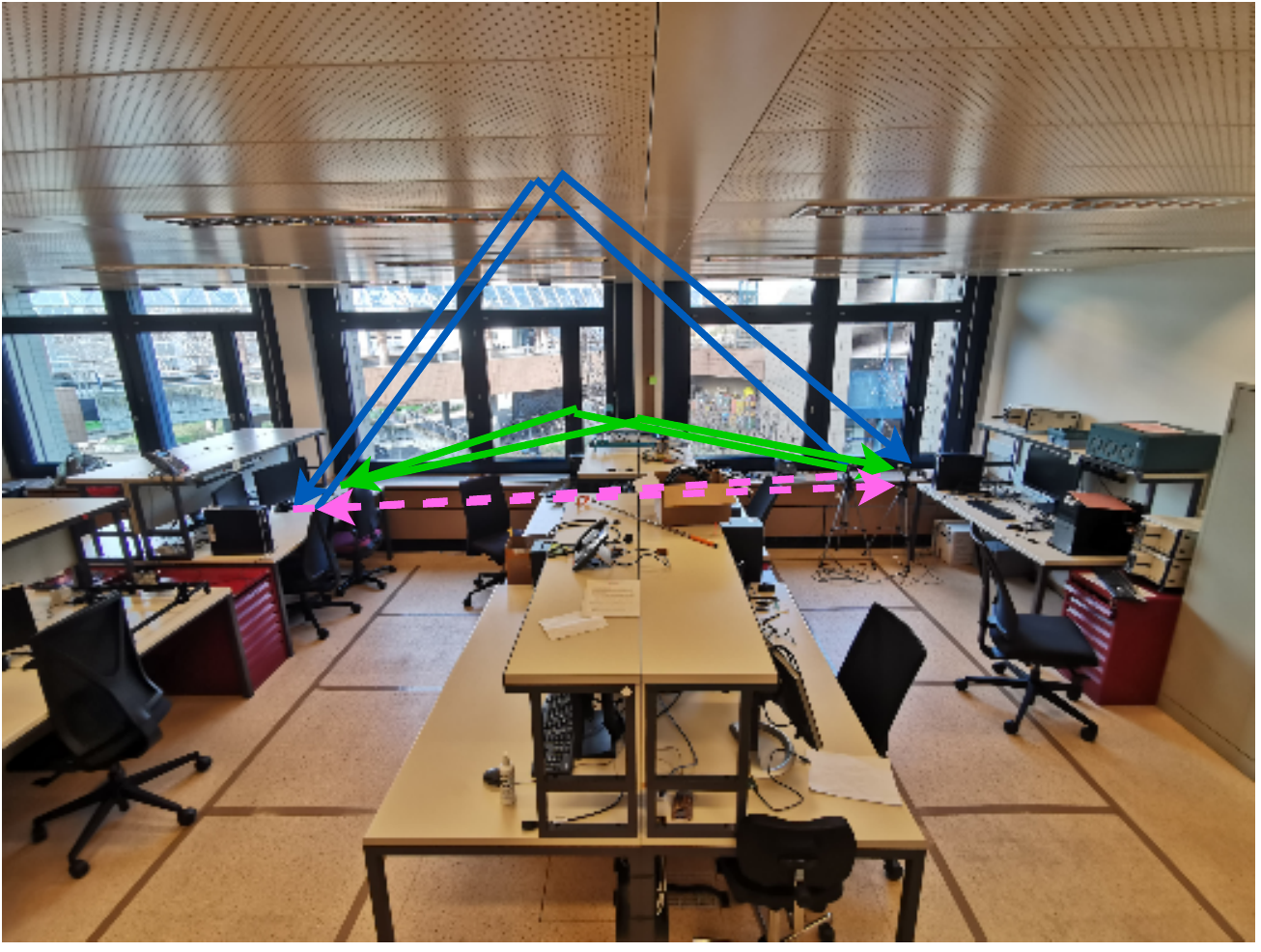}}
  \caption{Room layout and experiment setup}
  \label{fig:room} 
\end{figure}

\begin{table*}[t!] 
\centering
\caption{The median value of estimation error (in Hz) for each position ($0$ deg) with different displacements. 'Sel.' stands for BoI selection method and 'Fus.' stands for BoI fusion method.}
\label{tab:results}
\begin{tabular}{ccccccccccccc}
\toprule
             & \multicolumn{2}{c}{Position 1} & \multicolumn{2}{c}{Position 2} & \multicolumn{2}{c}{Position 3} & \multicolumn{2}{c}{Position 4} & \multicolumn{2}{c}{Position 5} & \multicolumn{2}{c}{Position 6} \\ \midrule
Displacement & Sel.           & Fus.          & Sel.           & Fus.          & Sel.           & Fus.          & Sel.           & Fus.          & Sel.           & Fus.          & Sel.           & Fus.          \\ \midrule
2 mm         & 0.110          & 0.002         & 0.171          & 0.192         & 0.204          & 0.186         & 0.084          & 0.111         & 0.139          & 0.175         & 0.050          & 0.001         \\
3 mm         & 0.077          & 0.002         & 0.098          & 0.002         & 0.144          & 0.233         & 0.079          & 0.102         & 0.119          & 0.130         & 0.002          & 0.002         \\
4 mm         & 0.027          & 0.001         & 0.019          & 0.001         & 0.176          & 0.180         & 0.157          & 0.174         & 0.110          & 0.120         & 0.002          & 0.001         \\
5 mm         & 0.003          & 0.001         & 0.003          & 0.001         & 0.137          & 0.167         & 0.108          & 0.039         & 0.075          & 0.056         & 0.003          & 0.001         \\
6 mm         & 0.002          & 0.001         & 0.001          & 0.001         & 0.110          & 0.078         & 0.071          & 0.108         & 0.091          & 0.062         & 0.001          & 0.001         \\
7 mm         & 0.001          & 0.001         & 0.001          & 0.001         & 0.080          & 0.106         & 0.102          & 0.064         & 0.137          & 0.001         & 0.001          & 0.001         \\
8 mm         & 0.001          & 0.001         & 0.001          & 0.001         & 0.067          & 0.061         & 0.125          & 0.041         & 0.054          & 0.001         & 0.001          & 0.001         \\ \bottomrule
\end{tabular}
\end{table*}

\begin{table*}[t!] 
\centering
\caption{The median value of estimation error (in Hz) for each position ($90$ deg) with different displacements. 'Sel.' stands for BoI selection method and 'Fus.' stands for BoI fusion method.}
\label{tab:results_90deg}
\begin{tabular}{ccccccccccccc}
\toprule
             & \multicolumn{2}{c}{Position 1} & \multicolumn{2}{c}{Position 2} & \multicolumn{2}{c}{Position 3} & \multicolumn{2}{c}{Position 4} & \multicolumn{2}{c}{Position 5} & \multicolumn{2}{c}{Position 6} \\ \midrule
Displacement & Sel.           & Fus.          & Sel.           & Fus.          & Sel.           & Fus.          & Sel.           & Fus.          & Sel.           & Fus.          & Sel.           & Fus.          \\ \midrule
2 mm         & 0.031          & 0.001         & 0.167          & 0.267         & 0.118          & 0.010         & 0.083          & 0.140         & 0.201          & 0.226         & 0.112          & 0.125         \\
3 mm         & 0.002          & 0.001         & 0.099          & 0.005         & 0.093          & 0.003         & 0.082          & 0.144         & 0.233          & 0.177         & 0.189          & 0.208         \\
4 mm         & 0.001          & 0.001         & 0.044          & 0.003         & 0.079          & 0.001         & 0.113          & 0.056         & 0.126          & 0.002         & 0.094          & 0.007         \\
5 mm         & 0.001          & 0.001         & 0.001          & 0.001         & 0.023          & 0.002         & 0.102          & 0.085         & 0.038          & 0.002         & 0.149          & 0.003         \\
6 mm         & 0.001          & 0.001         & 0.003          & 0.001         & 0.008          & 0.001         & 0.126          & 0.011         & 0.005          & 0.001         & 0.043          & 0.002         \\
7 mm         & 0.001          & 0.001         & 0.003          & 0.001         & 0.006          & 0.001         & 0.107          & 0.036         & 0.020          & 0.001         & 0.049          & 0.003         \\
8 mm         & 0.001          & 0.001         & 0.002          & 0.001         & 0.003          & 0.001         & 0.143          & 0.004         & 0.020          & 0.001         & 0.026          & 0.001         \\ \bottomrule
\end{tabular}
\end{table*}

\subsection{Band-of-Interest Fusion Method}
Instead of selecting one \gls{cir} bin as in~\cite{li_complexbeat_2021}, we propose to linearly combine multiple bins with a weight $\mathbf{w}$ in the \gls{cir} such that the signal energy within the \gls{boi} is maximized. 
The linear combination is calculated as $\mathbf{x} = \mathbf{H}\mathbf{w}$.
Based on~(\ref{equ:boi_equ}), the optimization problem can be written as 
\begin{align}
\max_\mathbf{w} \left( \mathbf{F}_I \mathbf{H} \mathbf{w} \right)^H \left( \mathbf{F}_I \mathbf{H} \mathbf{w} \right).
\end{align}
We also need to define a constraint, $\left( \mathbf{F} \mathbf{H} \mathbf{w} \right)^H \left( \mathbf{F} \mathbf{H} \mathbf{w} \right) = 1$, where $\mathbf{F}$ is the \gls{dft} matrix, to avoid the trivial all-infinite solution. 
By defining $\mathbf{A} := \mathbf{H}^H \mathbf{F}_I^H \mathbf{F}_I \mathbf{H}$ and defining $\mathbf{B} := \mathbf{H}^H \mathbf{F}^H \mathbf{F} \mathbf{H}$, the corresponding optimization problem is given by
\begin{align}
\arg \max_\mathbf{w} \quad &\mathbf{w}^H \mathbf{A} \mathbf{w}, \\
\text{s.t.} \quad & \mathbf{w}^H \mathbf{B} \mathbf{w} = 1.
\end{align}
We first write the stationary condition from the \gls{kkt} conditions~\cite{boyd_convex_2004} and obtain
\begin{align}
\mathbf{A} \mathbf{w}^* = \lambda \mathbf{B} \mathbf{w}^*,
\label{equ:generalized_eig_prob}
\end{align}
where $\lambda$ is the \gls{kkt} multiplier, which leads to a generalized eigenvalue problem~\cite{ghojogh_eigenvalue_2019}, denoted by $\left( \mathbf{A}, \mathbf{B} \right)$, and $\frac{\mathbf{w}^*}{\|\mathbf{w}^*\|}$ is an eigenvector of $\left( \mathbf{A}, \mathbf{B} \right)$.
We obtain the eigenvectors $\mathbf{w}_i'$ from the eigenvalue decomposition of $\left( \mathbf{A}, \mathbf{B} \right)$.
Afterwards, each $\mathbf{w}_i'$ is scaled to obtain one candidate $\mathbf{w}_i$ for the optimal solution based on the primal feasibility of the \gls{kkt} conditions with 
\begin{align}
\mathbf{w}_i = \left( \mathbf{w}_i^{'H}  \mathbf{B}  \mathbf{w}'_i \right)^{\frac{1}{2}} \mathbf{w}_i^{'}.
\end{align}
Finally, we select the $\mathbf{w}_i$ that leads to the largest $\mathbf{w}_i^H \mathbf{A} \mathbf{w}_i$ as the optimal solution $\mathbf{w}^*$.
The combined sequence based on the \gls{boi} is calculated by $\mathbf{x}^* = \mathbf{H}\mathbf{w}^*$.
This sequence is the input to the \gls{psd} peak detection block in the pipeline of our \gls{boi}-fusion method.

\section{Experimental Study}
In this section, we evaluate the performance of our \gls{boi} fusion method with experimental data and compare it with the \gls{boi} selection method.

\subsection{Experiment Setup}
The breathing displacement and breathing rate vary between individuals.
In order to test the performance of our method in different situations, we use a breathing emulator with a reflection plate of size $10$\,cm\,$\times$\,$20$\,cm to simulate the human chest movement during breathing.
We vary the moving speed and displacement of the breathing emulator to simulate different frequencies within the range of human breathing rates.
The breathing displacement varies from $2$\,mm to $8$\,mm~\cite{tong_respiratory-related_2019}, and the movement speed is in the range from $1.07$\,mm/s to $4.27$\,mm/s to emulate breathing rates from $0.1$\,Hz to $0.5$\,Hz, which leads to $74$ different breathing cases.
We also place the breathing emulator in different positions in the room, and rotate the breathing emulator $90$ degrees ($0$ degrees is towards north) such that the reflection plate is facing different directions to simulate different positions and orientations of humans.

The transmitter and the receiver are located in the corner of the computer room as shown in Fig.~\ref{fig:lab_raw}.
The breathing emulator is located on the desks in $6$ different positions.
At positions~$3$, $4$ and $5$, there is at least one obstacle (shelves, PC desktops, monitors, etc.) on the \gls{los} between the target and the transceivers.
For example, as shown in Fig.~\ref{fig:room_ray}, the pink \gls{los} path can be attenuated by the obstacles and there are potential paths with more reflections where we can extract periodic components.
While the breathing emulator is moving, we record \gls{cir} snapshots for $50$~seconds as one measurement for each position, each orientation, and each breathing case.
Two EVK1000~\cite{decawave_qurvo_nodate} working on \gls{uwb} channel $2$ are used as the transmitter and receiver.
The \gls{cir} is captured with a sampling rate of $19.3$\,Hz.\footnote{The sampling rate is programmed to $20$\,Hz. Because of packet loss and jitters, the achieved sampling rate is $19.3$\,Hz.}
Estimation is done in an $800$-snapshot sliding window, thus for each measurement, we obtain $50$\,s\,$\times$\,$19.3$\,Hz\,$-$\,$800= $\,$165$ breathing rate estimation windows.
Since there are jitters in the \gls{uwb} packets, we interpolate the data based on the received time stamps to remove the jitters in each window before estimation.
The resolution of the \gls{psd} peak detection is $0.001$\,Hz.
The estimates are compared with the ground truth breathing rate based on the moving speed and displacement of the breathing emulator.

\subsection{Results}
We compare the estimation accuracy of our \gls{boi} fusion method and the \gls{boi} selection method.
Apart from the absolute estimation error, we also want to have an insight into the \gls{psd} of the snapshot sequence.
When there is no distinct periodic component, the \gls{psd} of the snapshot sequence shows multiple peaks.
We want to check if the periodic component is distinct and the breathing rate estimate corresponds to the main peak in the \gls{psd} instead of having the largest peak that is coincidentally close to the ground truth breathing rate.
We, therefore, introduce another criterion: 
The confidence index indicates how likely the detected peak in the \gls{psd} of the snapshot sequence is the main peak.
The confidence index is calculated by the absolute difference between the first and the second peak in the \gls{psd} of the snapshot sequence.

\begin{figure}[t] 
  \centering
  \subfloat[Estimation error\label{fig:err_pos1_ns}]{%
        \includegraphics[width=0.48\linewidth]{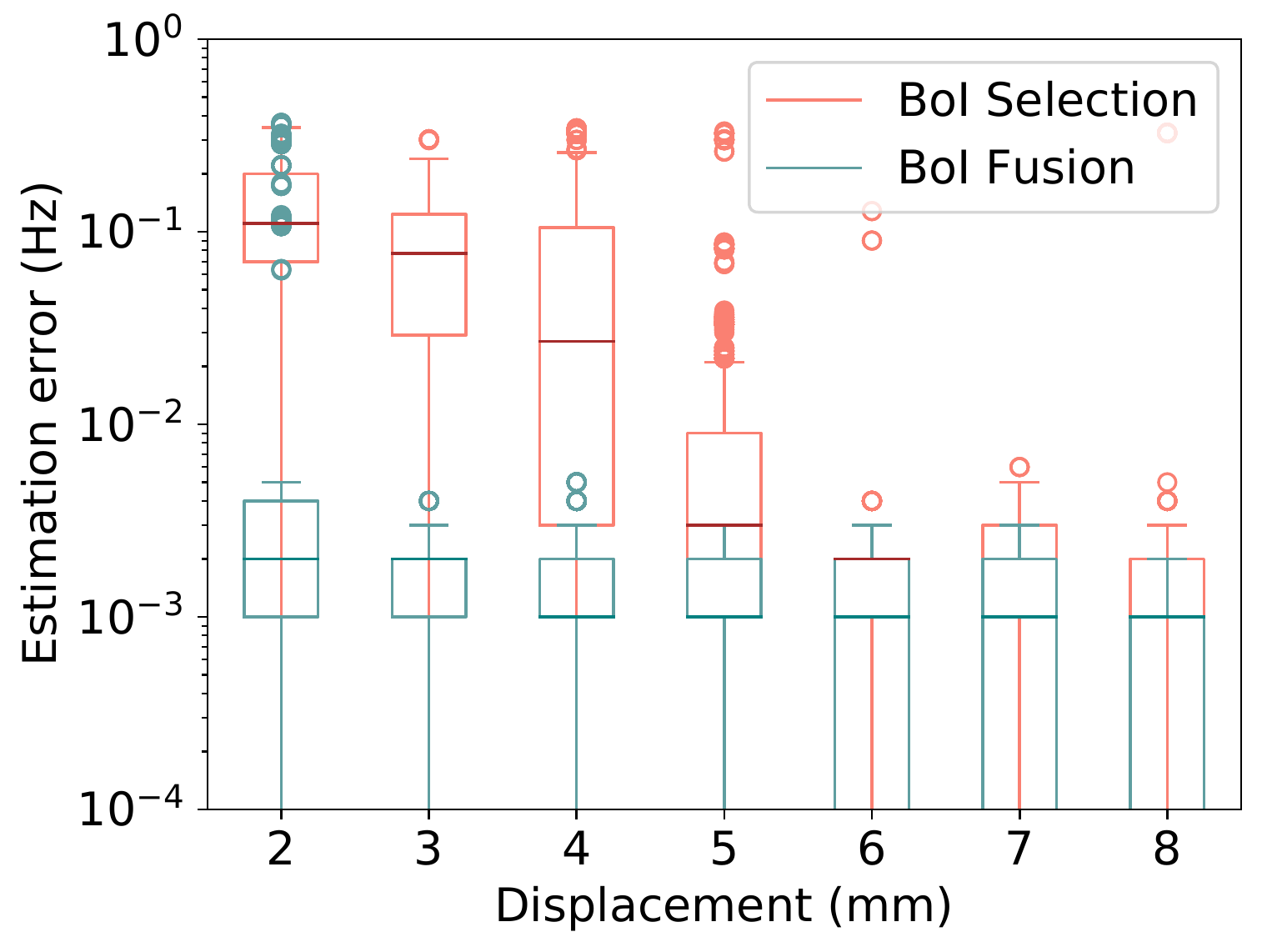}}
    \hfill
  \subfloat[Confidence index\label{fig:conf_pos1_ns}]{%
        \includegraphics[width=0.48\linewidth]{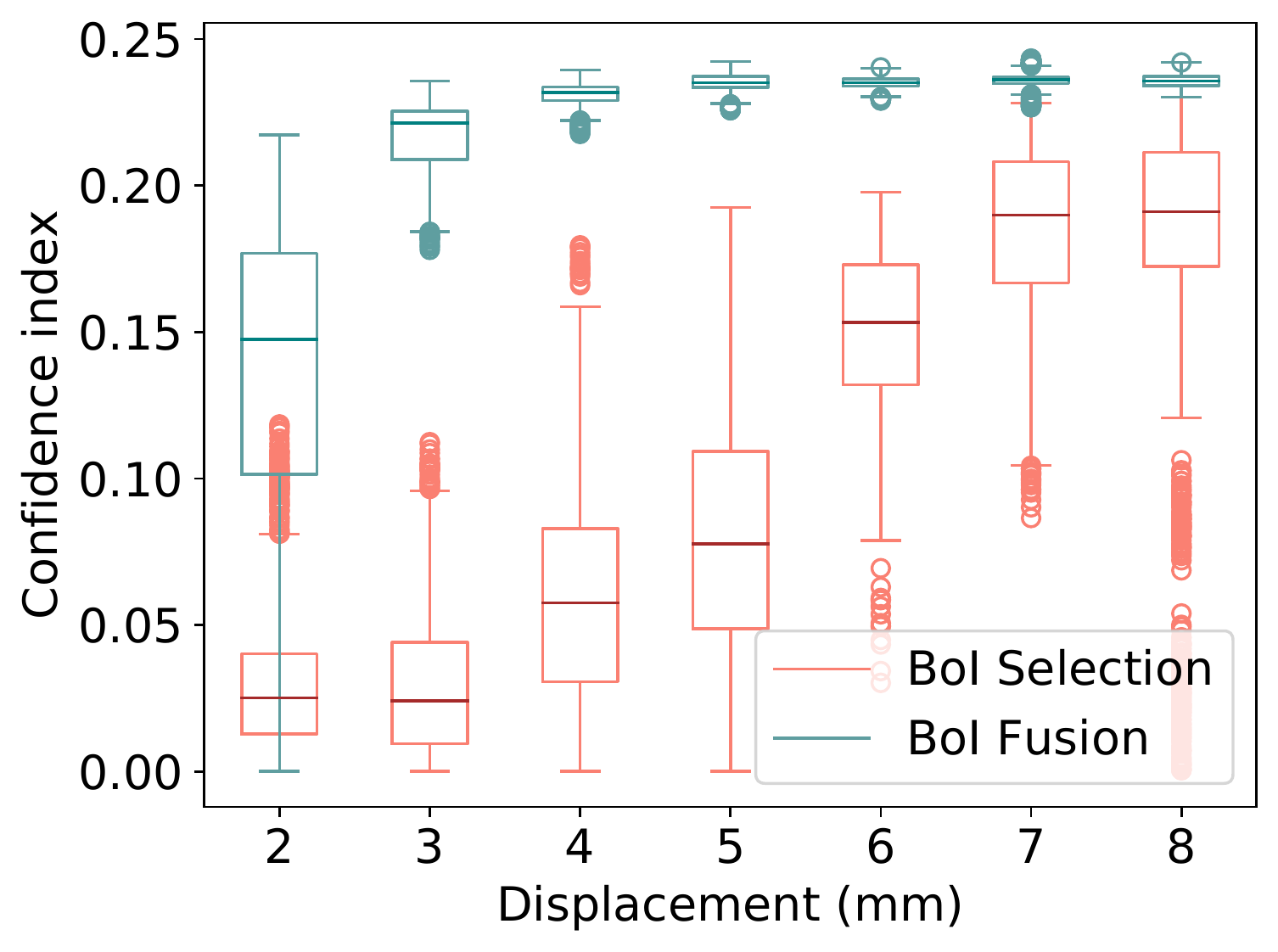}}
  \caption{Measurement at position $1$ (\gls{los} distance to \gls{uwb} devices $1.9$\,m), reflection plate displacement $2$\,mm to $8$\,mm, $0$\,deg}
  \label{fig:pos1_ew} 
\end{figure}

\begin{figure}[t] 
  \centering
  \subfloat[Estimation error\label{fig:err_pos5_ew}]{%
        \includegraphics[width=0.48\linewidth]{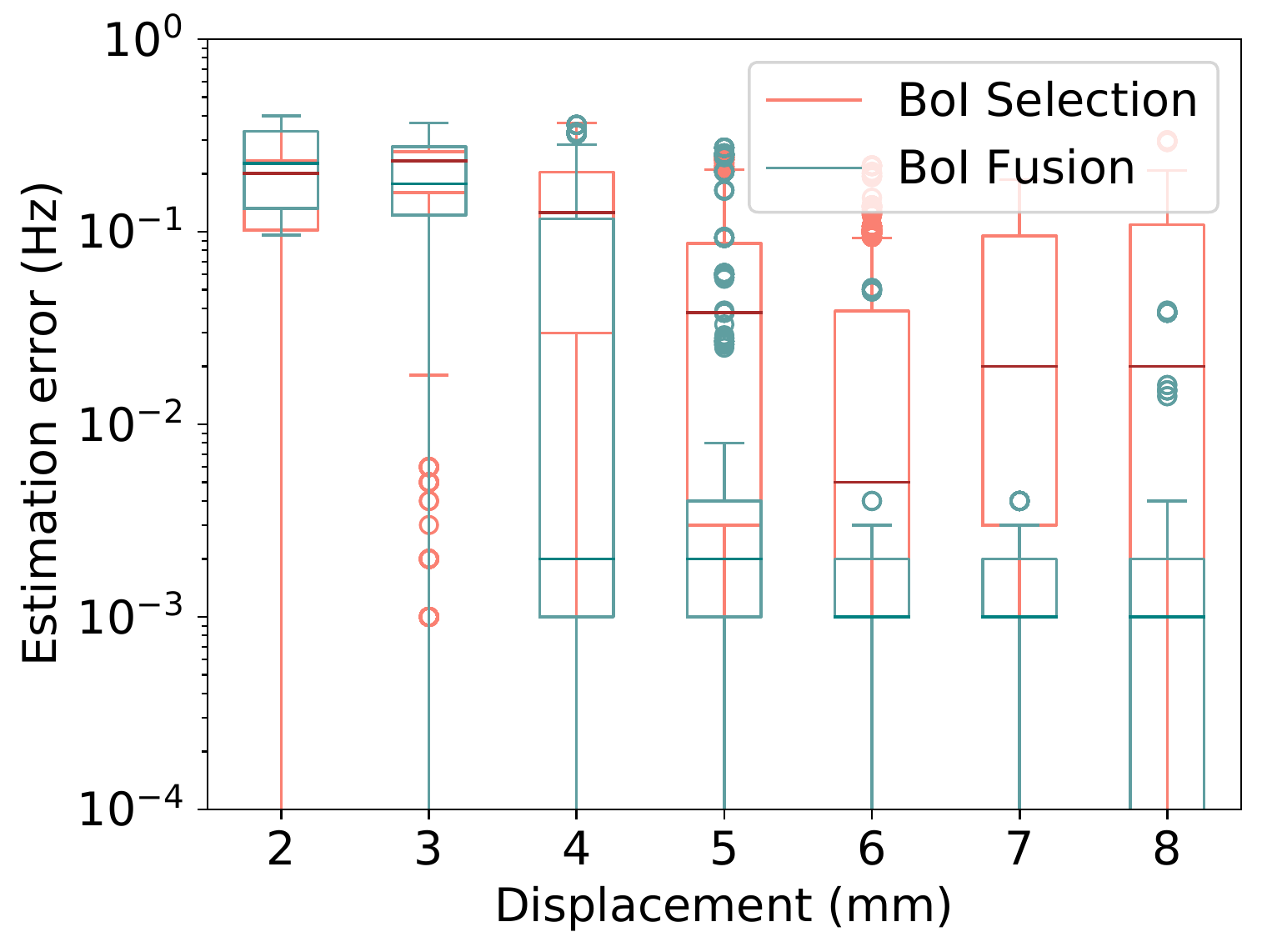}}
    \hfill
  \subfloat[Confidence index\label{fig:conf_pos5_ew}]{%
        \includegraphics[width=0.48\linewidth]{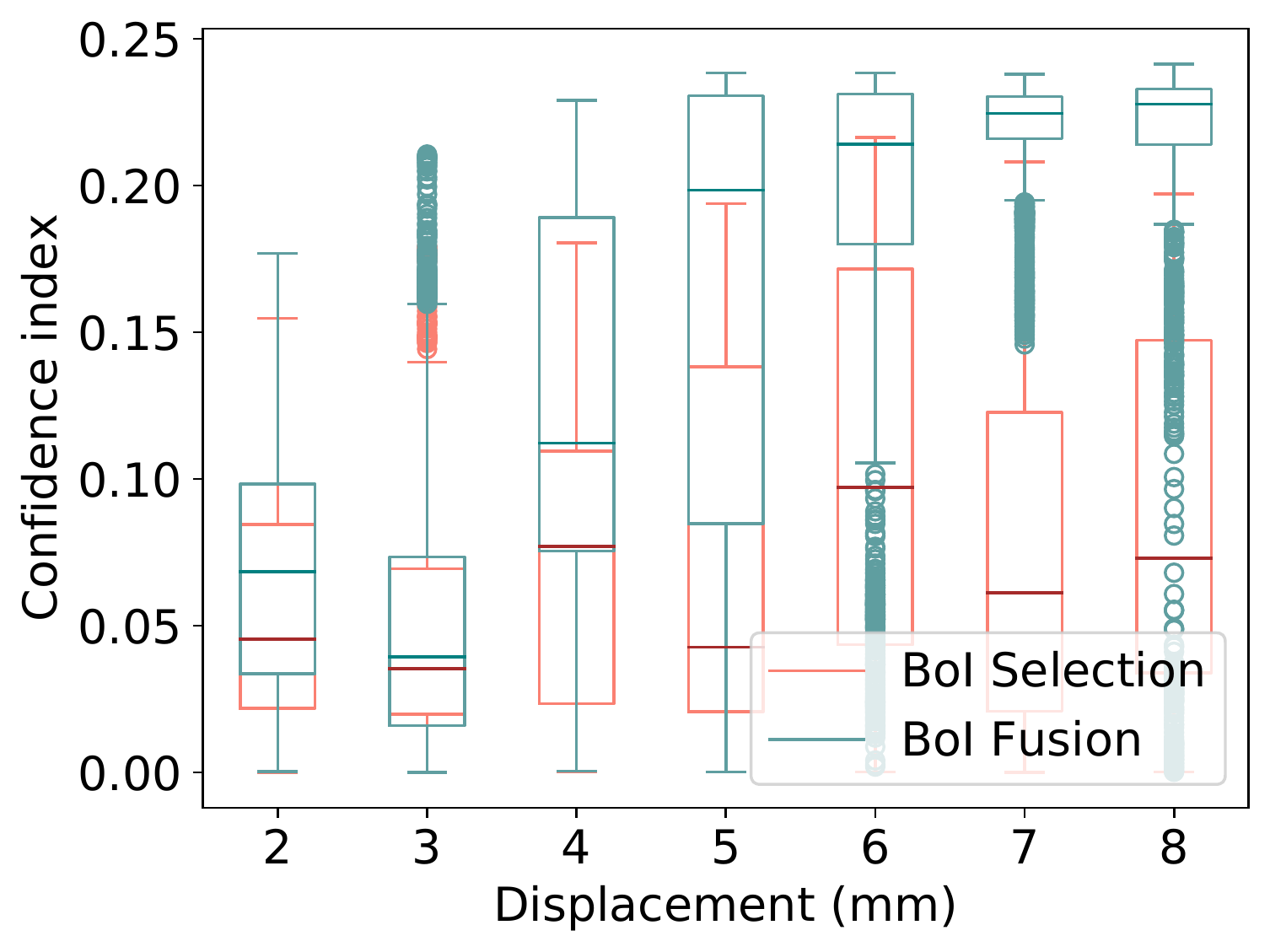}}
  \caption{Measurement at position 5 (\gls{nlos} distance to \gls{uwb} devices $5.5$\,m), reflection plate displacement $2$\,mm to $8$\,mm, $90$\,deg}
  \label{fig:pos5_ew} 
\end{figure}

\begin{figure}[t] 
  \centering
  \subfloat[Estimation error\label{fig:err_disp8_ns}]{%
        \includegraphics[width=0.48\linewidth]{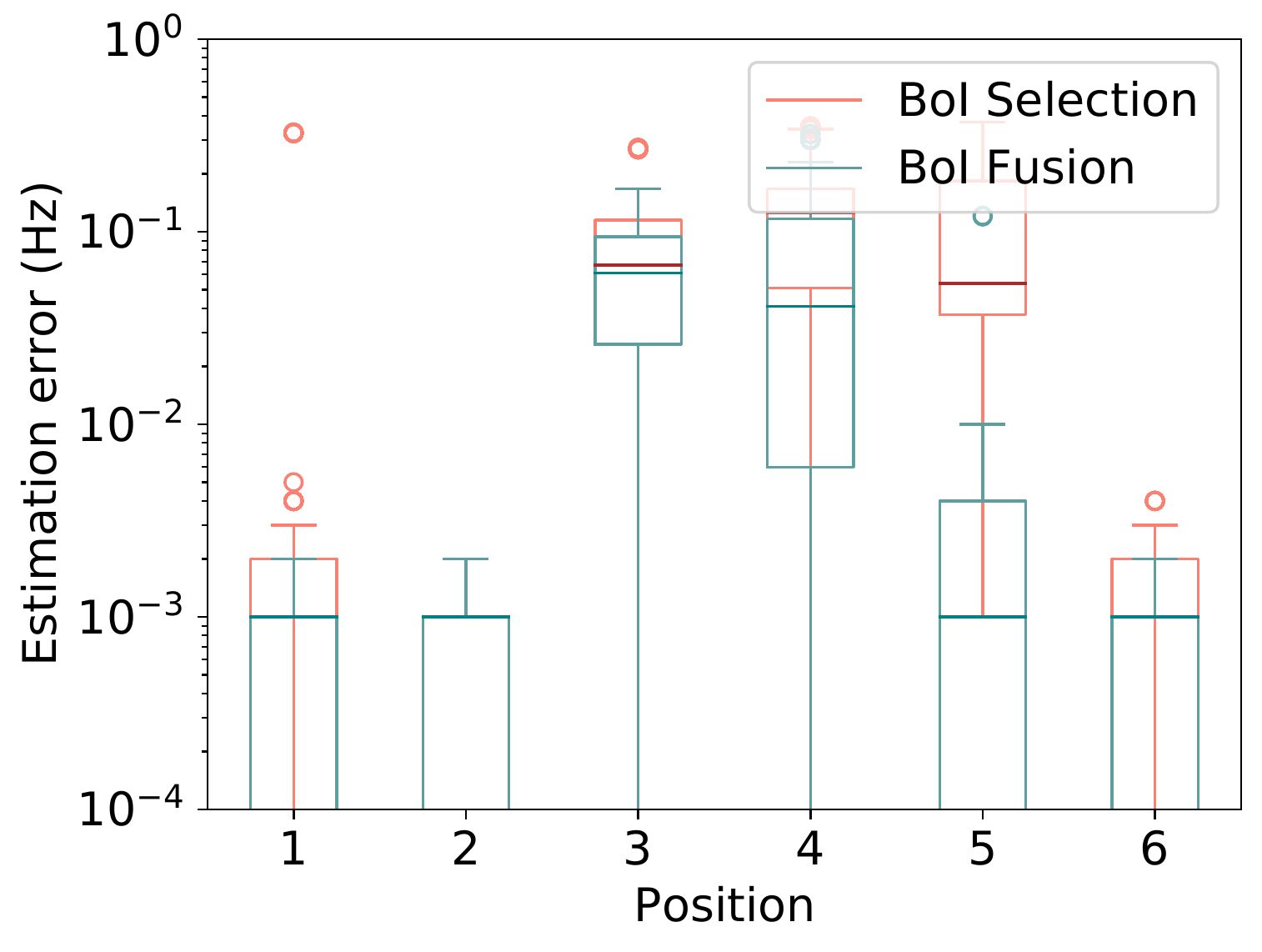}}
    \hfill
  \subfloat[Confidence index\label{fig:conf_disp8_ns}]{%
        \includegraphics[width=0.48\linewidth]{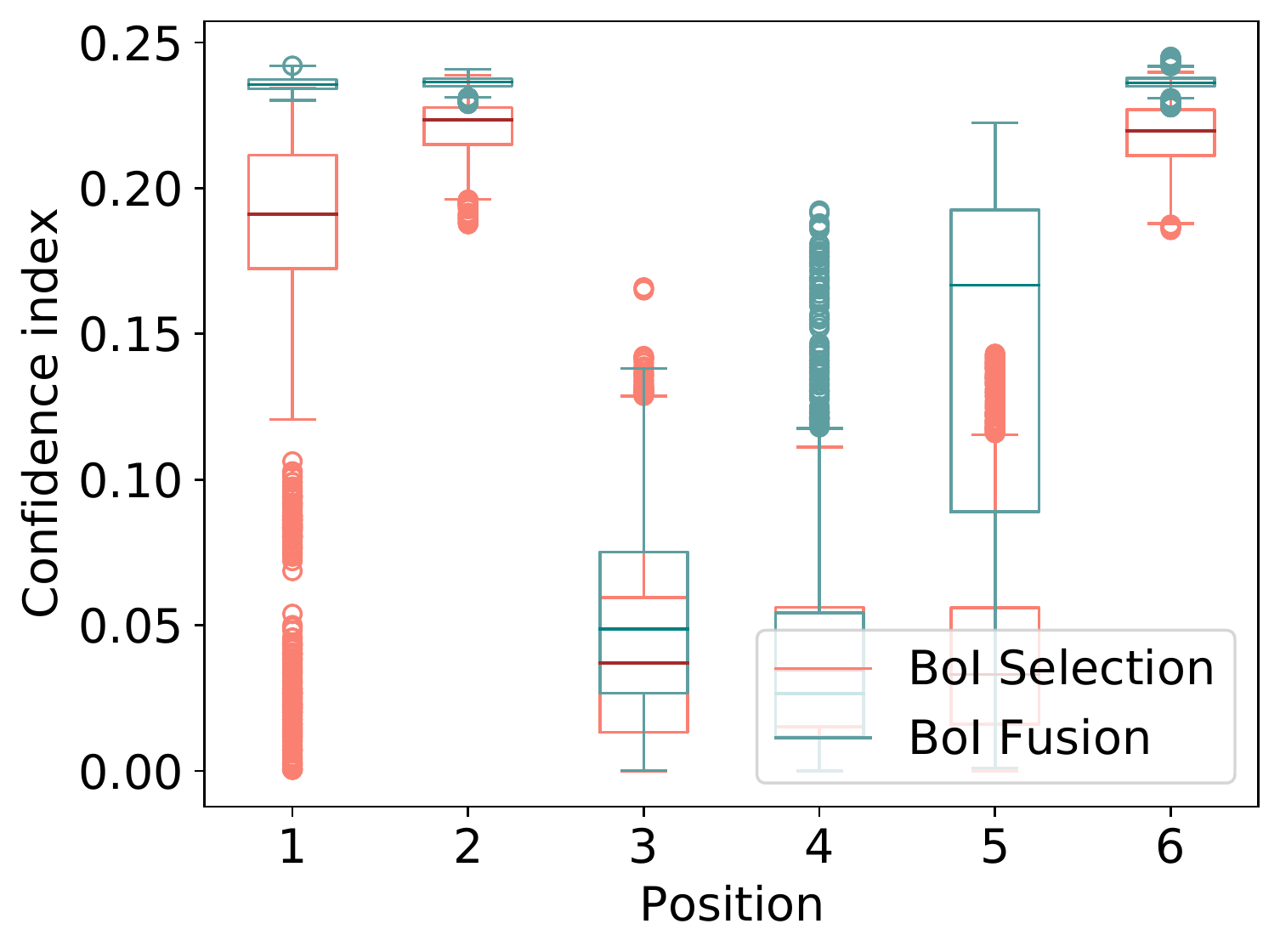}}
  \caption{Measurement with the displacement of $8$~mm, positions from $1$ to $6$, $0$\,deg}
  \label{fig:disp8_ns} 
\end{figure}

\begin{figure}[t] 
  \centering
  \subfloat[Estimation error\label{fig:err_disp8_ew}]{%
        \includegraphics[width=0.48\linewidth]{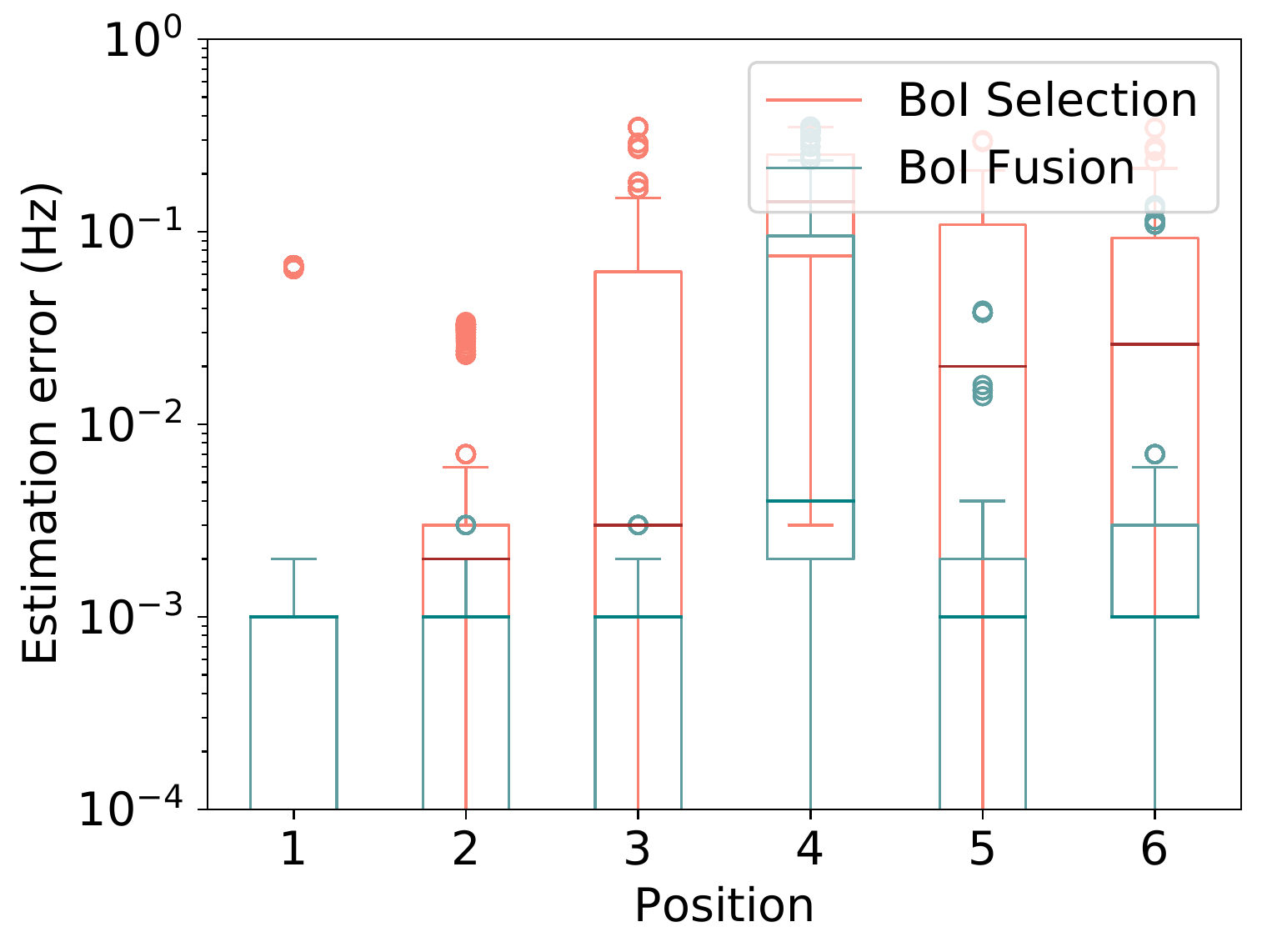}}
    \hfill
  \subfloat[Confidence index\label{fig:conf_disp8_ew}]{%
        \includegraphics[width=0.48\linewidth]{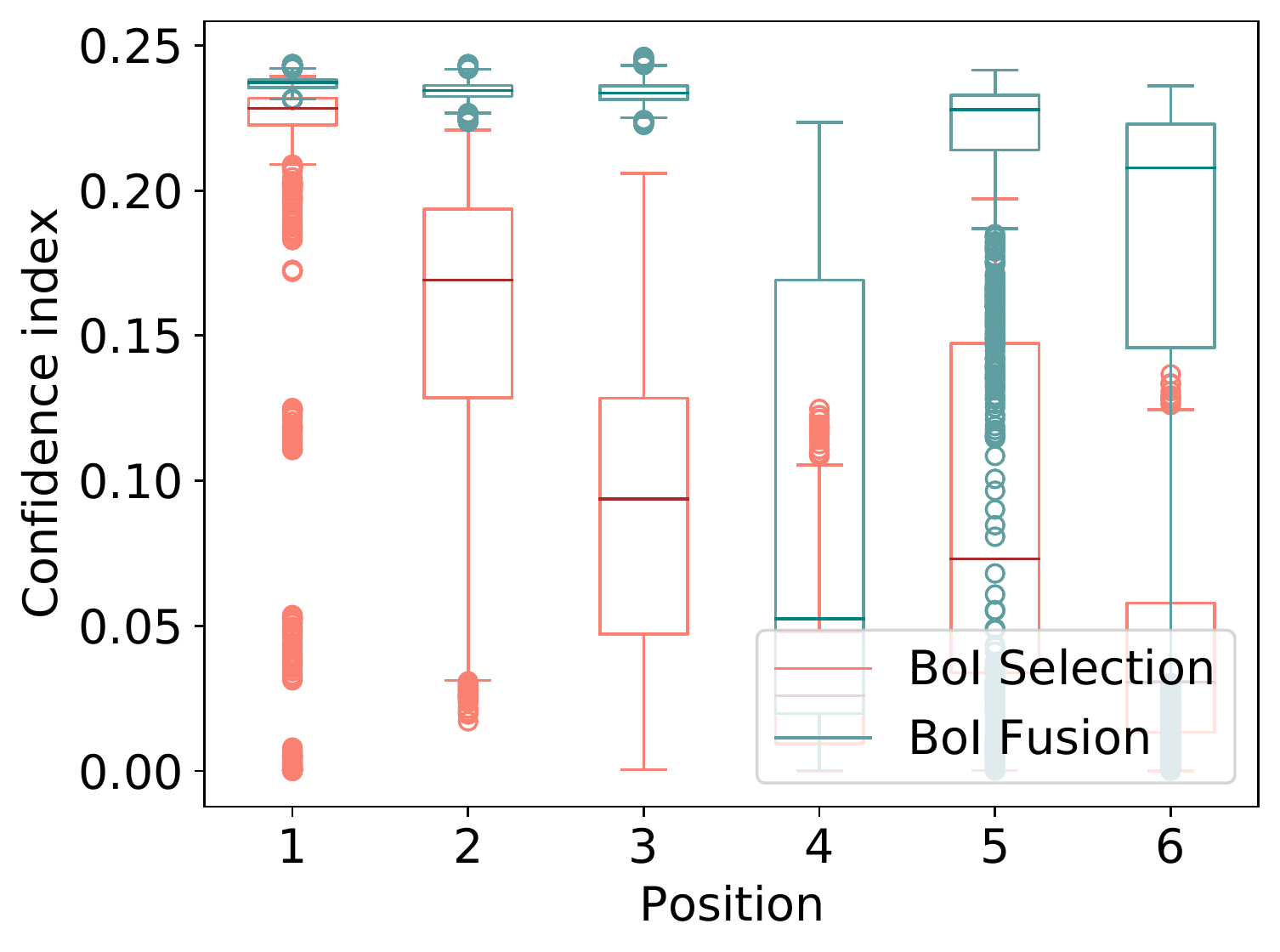}}
  \caption{Measurement with the displacement of $8$~mm, positions from $1$ to $6$, $90$\,deg}
  \label{fig:disp8_ew} 
\end{figure}

\subsubsection{Performance Overview}
The median value of the estimation error for the \gls{boi} selection method and the \gls{boi} fusion method is shown in Table~\ref{tab:results} ($0$~degrees) and Table~\ref{tab:results_90deg} ($90$~degrees).
For most of the scenarios, the \gls{boi} fusion method performs better than the \gls{boi} selection method.
However, we observed that the absolute performance and the performance improvement also depend on the displacement and the target position and orientation.
In the following, we evaluate the performance for these two parameters.

\subsubsection{Breathing Displacement}
We first observe the performance with different breathing displacements.
When the target is located in position $1$, i.e., the target is in the \gls{los} of the transceivers and is close to the transceivers, as shown in Fig.~\ref{fig:pos1_ew}, both methods show good performance when the displacement is larger than $5$\,mm.
However, when the displacement is small, even in the \gls{los} situation, the periodic component in a single \gls{cir} bin is very small.
Hence, the error of the \gls{boi} selection method is large.
On the contrary, our \gls{boi} fusion methods show good performance when the displacement is small.
For the confidence index, our method shows a median confidence index above $0.2$ for displacements larger than $2$\,mm, whereas the confidence index of the \gls{boi} selection method is lower and decreases when the displacement is smaller.
For position~$5$ which is further away from the transceiver and the \gls{los} is blocked by an obstacle, as shown in Fig.~\ref{fig:pos5_ew}, it is difficult for both methods to estimate the breathing rate when the displacement is smaller than $4$\,mm.
However, our \gls{boi} fusion method shows a higher confidence index and estimates the breathing rate with higher accuracy compared to the \gls{boi} selection method when the displacement increases.

\subsubsection{Object Position and Orientation}

We observe that the position and the orientation of the target influence the estimation performance.
As shown in Fig.~\ref{fig:disp8_ns}, for the largest displacement, it is more difficult to estimate the breathing rate in the \gls{nlos} positions, i.e., positions $3$, $4$ and $5$ for the \gls{boi} selection method.
However, the \gls{boi} fusion method shows better performance at positions $4$ and $5$.
When we rotate the breathing emulator for $90$ degrees as shown in Fig.~\ref{fig:disp8_ew}, it is less challenging for the \gls{boi} selection method at positions $3$ and $5$ and more difficult at position $6$.
However, our \gls{boi} fusion method still shows a smaller median estimation error compared to the \gls{boi} selection method at all positions.

\begin{table*}[] 
\centering
\caption{UWB Estimated breathing rate compared to breathing rate estimated by respiration monitor belt from Neulog at different positions and orientations}
\label{tab:real_human}
\begin{tabular}{ccccccccccccc}
\toprule
                     & \multicolumn{2}{c}{Position 1} & \multicolumn{2}{c}{Position 2} & \multicolumn{2}{c}{Position 3} & \multicolumn{2}{c}{Position 4} & \multicolumn{2}{c}{Position 5} & \multicolumn{2}{c}{Position 6} \\ \midrule
                     & 0 deg         & 90 deg         & 0 deg         & 90 deg         & 0 deg         & 90 deg         & 0 deg         & 90 deg         & 0 deg         & 90 deg         & 0 deg         & 90 deg         \\ \midrule
Neulog (Gound truth) & 0.300         & 0.323          & 0.291         & 0.284          & 0.268         & 0.285          & 0.278         & 0.264          & 0.258         & 0.266          & 0.291         & 0.326          \\
BoI Selection        & 0.308         & 0.325          & 0.296         & 0.285          & 0.112         & 0.285          & 0.113         & 0.264          & 0.145         & 0.260          & 0.288         & 0.326          \\
Boi Fusion           & 0.308         & 0.325          & 0.294         & 0.283          & 0.269         & 0.286          & 0.282         & 0.262          & 0.273         & 0.265          & 0.292         & 0.327          \\ \bottomrule
\end{tabular}
\end{table*}

\subsection{Validation with Human Breathing}
Finally, we test the performance with real human breathing in different positions and orientations.
The person is always sitting on the chairs for each position.
In $0$ degrees case, the person is facing north, and in the $90$ degrees case, the person is always facing his desk.
The person is wearing a respiration monitor belt and the breathing rate is recorded by Neulog devices~\cite{neulog_neulog_nodate} as ground truth.
We found that it is easier to estimate the breathing rate for humans compared to that of the breathing emulator for the most challenging positions $3$ and $4$ since the moving volume of real human breathing is larger than the reflection plate.
However, as shown in Table.~\ref{tab:real_human}, there are still challenging cases, e.g., position $4$ ($0$\,degrees), position $3$ ($0$\,degrees), and position $5$ ($0$\,degrees), where the estimated breathing rate from the \gls{boi} selection method is not close to the ground truth breathing rate.
For those cases, the proposed \gls{boi} fusion method can achieve  a better estimate.

\section{Conclusions}
In this paper, we showed that for breathing rate estimation, when there are obstacles between the target and the transceivers, and when the breathing displacement is small, it is difficult to extract the breathing rate based on only one \gls{cir} delay bin.
Moreover, the position and orientation of the target also influence the performance of breathing rate estimation.
To deal with those difficult cases, our \gls{boi} fusion method combines the bins that contain energy in the \gls{boi} in an effective manner and maximized the breathing periodic component and leads to a more accurate and reliable breathing rate estimate. 

\balance
\bstctlcite{IEEEexample:BSTcontrol}
\bibliographystyle{IEEEtran}

{\small\bibliography{23ICC_vital_sign_sensing}}

\end{document}